\documentclass[a4paper,11pt]{article}

\usepackage{jheppub} 
\usepackage{lineno}
\usepackage{booktabs}
\usepackage{comment}
\usepackage{subcaption}
\usepackage{graphicx}
\usepackage{float}
\usepackage{amsmath}
\usepackage{xspace}
\usepackage{amssymb}
\usepackage{booktabs}
\usepackage{multirow}
\usepackage{siunitx}
\DeclareSIUnit\barn{b}
\arxivnumber{2412.10141} 

\title{\boldmath Searching for long-lived dark scalars at the FCC-ee}

\author[a,1]{G. Ripellino\note{Corresponding author.}}
\author[b,c]{M. Vande Voorde}
\author[a]{A. Gall\'en}
\author[a]{R. Gonzalez Suarez}

\affiliation[a]{Department of Physics and Astronomy, Uppsala University,\\ L{\"a}gerhyggsv{\"a}gen 1, Uppsala, 752 37, Sweden}  
\affiliation[b]{Department of Physics, KTH Royal Institute of Technology,\\ Roslagstullbacken 21, Stockholm, 114 21, Sweden}
\affiliation[c]{Oskar Klein Center, AlbaNova University Centre, Stockholm, 106 91, Sweden}

\emailAdd{giulia.ripellino@cern.ch}

\abstract{
This paper investigates the search for long-lived dark scalars from exotic Higgs boson decays at the Future Circular Collider in its $e^+e^-$ stage, FCC-ee, considering an integrated luminosity of \SI{10.8}{\per\atto\barn} collected during the ZH run at a center-of-mass energy $\sqrt{s}=\SI{240}{\giga\electronvolt}$. The work considers $Zh$ events where the $Z$ boson decays leptonically and the Higgs boson $h$ decays into two long-lived dark scalars $s$ which further decay into bottom anti-bottom quark pairs. 
The analysis is performed using a parametrized simulation of the IDEA detector concept and targets dark scalar decays in the tracking volume, resulting in multiple displaced vertices in the final state. The sensitivity towards long-lived dark scalars at FCC-ee is estimated using an event selection requiring two opposite-charge, same-flavor leptons compatible with the $Z$ boson, and at least two displaced vertices in the final state. The selection is seen to efficiently remove the Standard Model background, while retaining sensitivity for dark scalar masses between $m_s=\SI{20}{\giga\electronvolt}$ and $m_s=\SI{60}{\giga\electronvolt}$ and mean proper lifetimes $c\tau$ between approximately \SI{10}{\milli\meter} and \SI{10}{\meter}. The results show that the search strategy has potential to probe Higgs to dark scalar branching ratios as low as \num{e-4} for a mean proper lifetime $c\tau\approx\SI{1}{\meter}$. The results provide the first sensitivity estimate for exotic Higgs decays at FCC-ee with the IDEA detector concept, using the common FCC framework.
}

\begin{document}

\maketitle
\flushbottom

\section{Introduction}~\label{section:introduction}
The discovery of the Higgs boson at CERN's Large Hadron Collider (LHC) in 2012 provided a key missing piece of the Standard Model (SM) of particle physics. Its properties have since then been under scrutiny at the LHC, but ultimate precision requires a dedicated future collider able to produce large amounts of Higgs bosons in a clean experimental environment. 
While precise measurements of the Higgs properties have the potential to give indirect insights on physics beyond the SM (BSM), many theories point to the Higgs boson as a possible portal to new physics, with exotic Higgs boson decays into new light particles being the primary phenomenological consequence and means of discovery. The direct search for exotic Higgs boson decays is therefore an important complement to the precision program at future Higgs factories.

Exotic Higgs boson decays into promptly decaying particles have been extensively explored at the LHC~\cite{Curtin:2013fra,LHCHiggsCrossSectionWorkingGroup:2016ypw} and in the context of future Higgs factories~\cite{Liu:2016zki}. However, new physics may also manifest itself in exotic Higgs boson decays into particles with macroscopic mean proper lifetimes ($c\tau \gtrsim \SI{100}{\micro\meter}$), known as long-lived particles (LLPs). Such signals are predicted by a variety of BSM scenarios and feature unconventional experimental signatures which call for strategies outside the scope of typical collider analyses.

The Future Circular Collider (FCC) is a proposed future collider that could succeed the LHC at CERN following the priorities set by the 2020 Update of the European Strategy for Particle Physics~\cite{CERN-ESU-015}. The FCC program is designed to proceed in two stages. The first stage (FCC-ee) is a high-luminosity, high-precision, electron-positron collider operating as an electroweak, top quark and, especially, Higgs boson factory~\cite{FCC-CDR2}. Even though the FCC-ee will be primarily a high precision exploration tool, it also holds potential to directly discovering new physics~\cite{Curtin:2013fra}. The FCC-ee ZH run at center-of-mass energy $\sqrt{s}=\SI{240}{\giga\electronvolt}$ is expected to produce about two million Higgs bosons in a clean experimental environment, providing excellent opportunities for direct searches for LLPs from exotic Higgs decays. Previous studies suggest that the FCC-ee sensitivity to LLPs from exotic Higgs decays is broadly competitive with that of the LHC after its high-luminosity phase (HL-LHC)~\cite{Alipour-Fard:2018lsf}. In this paper we present the first study of exotic Higgs decays into long-lived scalars based on the IDEA detector concept, using the common FCC framework.

The exotic Higgs decays into long-lived scalars are realized in a scalar extension of the SM where a new real scalar field $S$ interacts with the Higgs field doublet $H$ via the portal term $\frac{\kappa}{2} S^2 |H|^2$. This simple construction is intimately connected to many BSM theories, such as Folded SUSY~\cite{Cohen:2015gaa, Burdman:2006tz} and Quirky Little Higgs~\cite{Cai:2008au}, that see the Higgs as a natural connection to still hidden sectors of new physics. 
The coupling $\kappa$ generates a mixing between the physical Higgs boson $h$ and the new dark scalar $s$ with a small mixing angle $\sin\theta$. As a consequence, the dark scalar can be produced in pairs in exotic decays of the Higgs boson and may also decay back into SM particles. 
The lifetime of the dark scalar is controlled by the mixing angle, with values $\sin\theta \lesssim \num{e-5}$ resulting in lifetimes $c\tau \gtrsim \mathcal{O}(\SI{1}{\milli\meter})$.

\begin{figure}[ht]\centering
\includegraphics[width=0.7\textwidth]{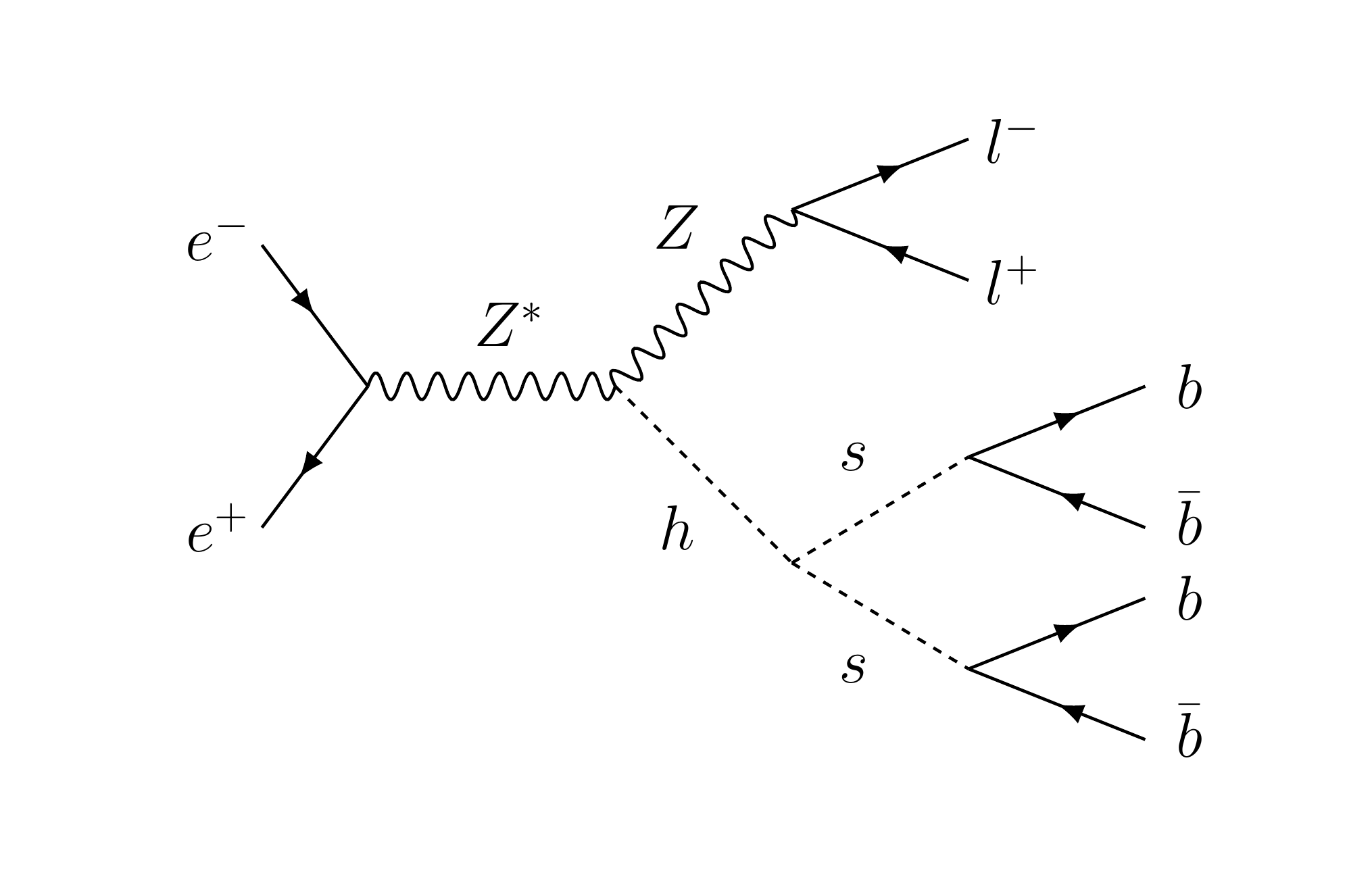}
    \caption{Feynman diagram of the complete signal process $e^+e^- \to Zh, Z\to l^+l^-, h\to ss \to b\bar{b}b\bar{b}$.}.
    \label{fig:feynman}    
\end{figure}

This work considers Higgs boson production in association with a $Z$ boson in $e^+e^-$ collisions at $\sqrt{s}=\SI{240}{\giga\electronvolt}$, considering the predicted integrated luminosity of \SI{10.8}{\per\atto\barn} of the FCC-ee ZH run~\cite{lumiFCC}. The long-lived scalars are produced in pairs through decays of the Higgs boson and then decay exclusively into $b\bar{b}$, as shown in figure~\ref{fig:feynman}. The analysis exploits the performance of the Innovative Detector for Electron-positron Accelerators (IDEA)~\cite{Antonello:2020tzq} and targets long-lived scalar decays within the tracking volume, resulting in final states with displaced vertices (DVs) reconstructed from inner-detector tracks. The associated Higgs boson production topology is also probed in the analysis, exploiting the clean signature from $Z$ boson decays into electrons or muons. Events selected for the analysis are thus required to contain at least two DVs with high mass and high track multiplicity relative to DVs from SM processes, as well as an $e^+e^-$ or $\mu^+\mu^-$ pair with an invariant mass consistent with the $Z$ boson mass. The event selection is seen to effectively remove all background from SM processes. 

Several searches for $h\to ss \to b\bar{b}b\bar{b}$ decays have been performed at the LHC, optimized for different regimes of $c\tau$. Searches performed by the ATLAS and CMS experiments using inner-detector information exclude $h\to ss$ branching ratios larger than \SI{10}{\percent} in the range $\SI{2}{\milli\meter} \lesssim c\tau \lesssim \SI{200}{\milli\meter}$ for $m_s$ between \SI{15}{\giga\electronvolt} and \SI{60}{\giga\electronvolt}~\cite{ATLAS:2024qoo,CMS:2024xzb}. These limits can be significantly extended when considering the full combined LHC and HL-LHC dataset. Projections suggest that searches based on inner-detector DVs have potential to probe branching ratios below \num{e-4} in a lifetime regime $\SI{10}{\micro\meter} \lesssim c\tau \lesssim \SI{1}{\meter}$~\cite{Curtin:2015fna}. In this paper we show that a similar search strategy at the FCC-ee also has potential to probe branching ratios below \num{e-4} for lifetimes around \SI{1}{\meter}.

\section{Signal model definition and simulation}~\label{section:signal}
This work considers an extension of the SM by a new real scalar field $S$ which interacts with the SM Higgs boson doublet $H$. The minimal renormalizable Lagrangian describing the scalar system is given by
\begin{equation}
\mathcal{L}_\text{scalar} = \mathcal{L}_\text{kin} +
\frac{\mu_s^2}{2} S^2 - \frac{\lambda_s}{4!} S^4 - 
\frac{\kappa}{2} S^2 |H|^2 + 
\mu^2 \left| H \right|^2 - \lambda \left| H \right|^4 \,,
\end{equation}
where a discrete symmetry taking $S \to -S$ is assumed. Depending on the choice of couplings, the two fields may acquire non-zero vacuum expectation values, and the two scalar states, corresponding to the SM Higgs boson $h$ and the dark scalar $s$, mix with a small mixing angle $\sin\theta$. The mixing allows for decays of the SM Higgs boson into pairs of dark scalars, as well as decays of the dark scalar back into SM states. The decay width of the dark scalar into SM states is equal to that of the SM Higgs boson with the mass of the dark scalar, reduced by a factor $\sin^2\theta$, according to 
\begin{equation} \label{eq:scalar_width}
\Gamma(s \to X_\text{SM}X_\text{SM}) = \sin^2 \theta \cdot \Gamma(h(m_s) \to X_\text{SM}X_\text{SM}) \,. 
\end{equation}
The mixing suppression $\sin^2\theta$ is common to all partial widths, and therefore does not affect the branching ratios if $s$ only decays to SM particles. For dark scalar masses above \SI{10}{\giga\electronvolt}, the decay into $b\bar{b}$ is the dominant decay channel, with a branching ratio of $0.9$~\cite{Curtin:2013fra}.

The signal process $e^+e^- \to Zh, \, Z\to l^+l^-, \, h\to ss \to b\bar{b}b\bar{b}$ is generated based on the Hidden Abelian Higgs Model (HAHM)~\cite{Curtin:2013fra,Curtin:2014cca}. The HAHM model includes a dark photon mediator, which here is decoupled by setting its mixing with the SM to zero, reducing the model to the simple scalar extension of the SM. Two free parameters then remain; the mass of the dark scalar $m_s$ and the coupling constant $\kappa$. 

Signal samples are generated using the HAHM\_MG5Model\_v3 model~\cite{hahm_gitlab} imported to MadGraph5\_aMC@NLO~v3.2.0~\cite{Stelzer:1994ta,Alwall:2014hca} to simulate parton-level $e^+e^-$ collisions at $\sqrt{s}=\SI{240}{\giga\electronvolt}$ at leading order, and are propagated to Pythia~v8.303~\cite{PYTHIA8} to simulate parton showering and hadronization. A grid of signal samples is generated with dark scalar masses $m_s$ between \SI{20}{\giga\electronvolt} and \SI{60}{\giga\electronvolt}, and mixing angles $\sin\theta$ between \num{1e-7} and \num{1e-5}. 
For each signal sample, the width of the dark scalar is fixed in the simulation to the value corresponding to the considered mixing angle and dark scalar mass, according to eq.~\ref{eq:scalar_width} using the lowest order expression for the Higgs width.
The coupling constant is fixed to $\kappa = 0.0007$, such that the branching ratio $\text{BR}(h \to ss)$ is below \num{2e-3} for all signal points. This leads to a modification of the SM Higgs width well below the projected precision of \SI{1}{\percent} for the width when considering the full dataset from the HL-LHC~\cite{Cepeda:2019klc}. 

Each signal sample is generated with \num{10000} events and is normalized to match the number expected from the FCC-ee ZH run according to
\begin{equation} \label{eq:nevt}
    N_\text{evt} = N_{Zh} \cdot \text{BR}(Z \to l^+ l^-) \cdot \text{BR}(h \to ss) \cdot \text{BR} (s \to b \bar{b})^2,
\end{equation}
with $N_{Zh}=\num{2.2e6}$ being the predicted total number of $Zh$ events~\cite{lumiFCC}. The branching ratio for the $Z$ boson decay is set to the SM value and the branching ratio for the dark scalar decay to 0.9~\cite{Curtin:2013fra}. The branching ratio of the Higgs boson decay to the dark scalar is computed as
\begin{equation} \label{eq:higgsBRtoscalar}
    \text{BR}(h \to ss) = \frac{\Gamma(h \to ss)}{\Gamma(h \to \text{SM}) + \Gamma(h \to ss)},
\end{equation}
where $\Gamma(h \to \text{SM})$ is the SM width and $\Gamma(h\to ss)$ is computed analytically at lowest order according to
\begin{equation} \label{eq:higgstoscalar_width}
    \Gamma(h \to ss) = \frac{\kappa^2 v_h^2}{32\pi m_h} \frac{\left( m_h^2+2m_s^2 \right)^2}{\left( m_h^2 -m_s^2 \right)^2} \sqrt{1 - 4 \frac{m_s^2}{m_h^2}} \,,
\end{equation}
with $v_h$ being the Higgs vacuum expectation value and $m_h$ and $m_s$  the masses of the Higgs boson and the dark scalar~\cite{hahm_gitlab_readme}. 

\begin{figure}[ht]\centering
  \subfloat[]{\includegraphics[width=0.45\textwidth]{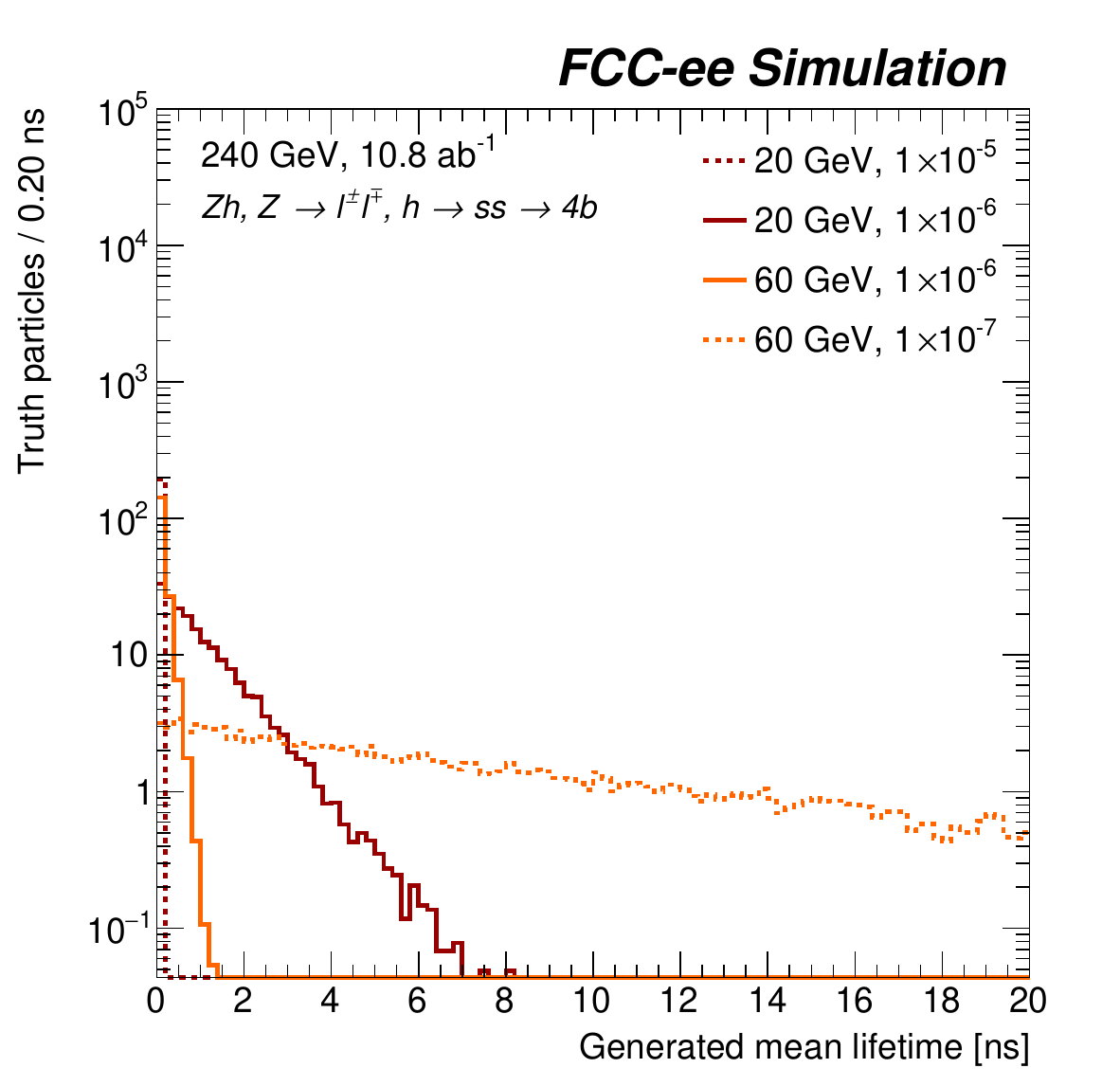}\label{fig:lifetime}}
  \subfloat[]{\includegraphics[width=0.45\textwidth]{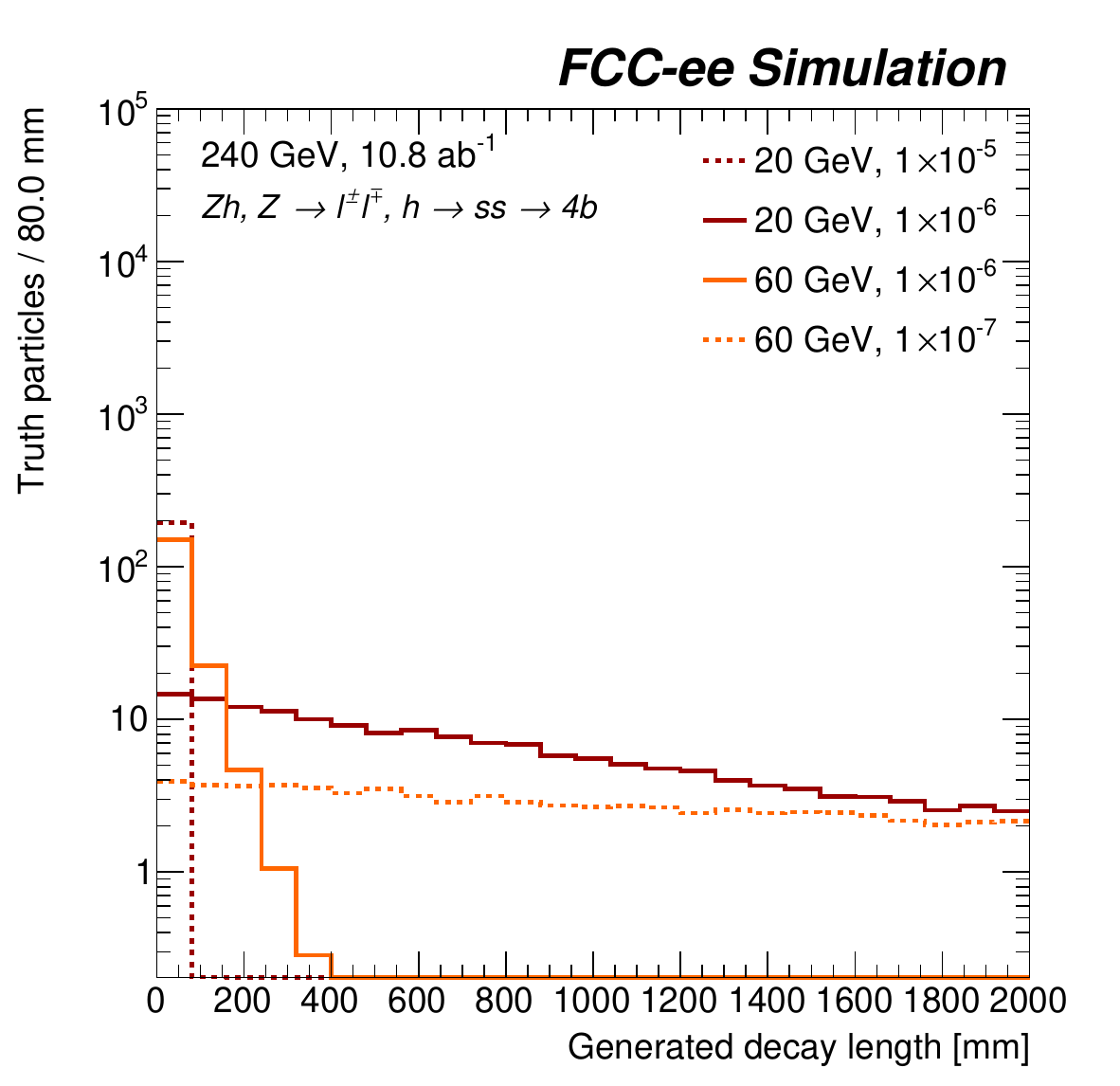}\label{fig:decaylength}}
 \caption{Generated dark scalar \protect\subref{fig:lifetime} lifetime and \protect\subref{fig:decaylength} decay length distributions for a selection of signal samples.}
    \label{fig:genlevel}
\end{figure}

The dark scalar lifetime and decay length distributions are shown for four representative signal samples in figure~\ref{fig:genlevel}. The lifetime of the dark scalar depends on the dark scalar mass and the mixing angle $\sin\theta$, with the smallest mixing angles and masses resulting in the longest lifetimes. All generated signal points are summarized in table~\ref{tab:grid}, together with the mean lifetime $c\tau$ computed from the width, and the branching ratio as computed from eq.~\ref{eq:higgsBRtoscalar}.

\begin{table}[ht]
\centering
\resizebox{0.47\textwidth}{!}{%
\begin{tabular}{@{}crr@{}}
\toprule
Sample ($m_s$, $\sin\theta$)& $c\tau$ [mm] & $\text{BR}(h \to ss)$ \\ \midrule \midrule 
\SI{20}{\giga\electronvolt}, \num{1e-5} & \num{3.4} & \num{8.1e-4} \\ 
\SI{20}{\giga\electronvolt}, \num{3e-6} & \num{38} & \num{8.1e-4} \\
\SI{20}{\giga\electronvolt}, \num{1e-6} & \num{340} & \num{8.1e-4} \\
\SI{20}{\giga\electronvolt}, \num{1e-7} & \num{34000} & \num{8.1e-4} \\ \midrule
\SI{40}{\giga\electronvolt}, \num{1e-5} & \num{1.4} & \num{10.2e-4} \\
\SI{40}{\giga\electronvolt}, \num{1e-6} & \num{140} & \num{10.2e-4} \\
\SI{40}{\giga\electronvolt}, \num{1e-7} & \num{14000} & \num{10.2e-4} \\ \midrule
\SI{50}{\giga\electronvolt}, \num{3e-6} & \num{12} & \num{10.9e-4} \\
\SI{50}{\giga\electronvolt}, \num{1e-6} & \num{110} & \num{10.9e-4} \\
\SI{50}{\giga\electronvolt}, \num{3e-7} & \num{1200} & \num{10.9e-4} \\ \midrule
\SI{60}{\giga\electronvolt}, \num{1e-5} & \num{0.9} & \num{7.4e-4} \\
\SI{60}{\giga\electronvolt}, \num{1e-6} & \num{88} & \num{7.4e-4} \\
\SI{60}{\giga\electronvolt}, \num{1e-7} & \num{8800} & \num{7.4e-4} \\
\bottomrule
\end{tabular}
}
\caption{List of generated signal points, including the dark scalar mass $m_s$, the mixing parameter $\sin\theta$, the mean lifetime $c\tau$, and the Higgs to dark scalar branching ratio $\text{BR}(h \to ss)$. The values of $m_s$ and $\sin\theta$ are fixed in the simulation, while the values of $c\tau$ and $\text{BR}(h \to ss)$ are computed analytically.}
\label{tab:grid}
\end{table}

\section{Background simulation}~\label{section:background}
Sources of SM background come from production of $Zh$, $ZZ$ and $WW$ at $\sqrt{s}=240$~GeV, with subsequent heavy-flavor and tau decays resulting in displaced SM vertices. The production of $Zh$ and $ZZ$ with a leptonic $Z$-boson decay dominate while all other background sources are effectively suppressed by requiring the presence of two opposite-sign leptons compatible with the $Z$-boson mass. All $Zh$, $ZZ$ and $WW$ processes with SM final states are taken into account when estimating the background. Background contributions from $e^{+}e^{-} \rightarrow e^{+}e^{-}$ and $e^{+}e^{-} \rightarrow \mu^{+}\mu^{-}$ have been assessed in dedicated simulated samples, and are found to be negligible. Beam-induced backgrounds are also expected to be negligible in the analysis~\cite{boscolo_2025_p44x1-18z28}.

Background samples are centrally produced within the FCC-ee ``Winter 2023" campaign~\cite{winter20233}, using Pythia~8.306~\cite{PYTHIA8} for $ZZ$ and $WW$, and Whizard~3.0.3~\cite{Kilian2011-ul,moretti2001omega} interfaced with Pythia~6.4.28~\cite{Sjostrand:2006za} for $Zh$ production. A total of \num{373e6} events are simulated for $WW$ production, \num{56e6} for $ZZ$, and \num{33e6} for $Zh$ production. The background samples are normalized to the expected integrated luminosity of \SI{10.8}{\per\atto\barn}.

\section{Event reconstruction}~\label{section:reconstruction}
The detector response is simulated with Delphes~v3.5.1pre05~\cite{deFavereau:2013fsa}, using the IDEA detector card. The detector concept comprises of a silicon pixel vertex detector; a large-volume, light short-drift wire chamber; a thin, low-mass superconducting solenoid coil; a pre-shower detector; a dual-readout calorimeter; and muon chambers within the magnet return yoke. The analysis is based on tracks, vertices and leptons reconstructed in Delphes with the parametrized IDEA detector resolution. These Delphes objects are converted into the EDM4HEP format~\cite{EDM4hep}, using the k4SimDelphes project~\cite{k4simdelphes}, and are analyzed with the FCCAnalyses~\cite{FCCAnalyses} framework. 

\subsection{Track and primary vertex reconstruction}
Tracks and primary vertices are reconstructed in Delphes based on the geometry of the IDEA tracking system. The vertex detector is simulated as five cylindrical barrel layers with radius between \SI{1.2}{\centi\meter}--\SI{31.5}{\centi\meter}, and three endcap disks in each forward region. The layers provide two-dimensional measurement points with a resolution between \SI{3}{\micro\meter}--\SI{7}{\micro\meter}. The drift chamber is modeled as 112 co-axial layers providing a single measurement point with a resolution of \SI{100}{\micro\meter}. It has a length of \SI{4}{\meter}, covers a radial region between \SI{34}{\centi\meter}--\SI{2}{\meter}, and extends up to a pseudorapidity of approximately $\eta=2.5$. 

The Delphes simulation software accounts for the finite detector resolution and the multiple scattering in each tracking layer. It provides parameterized track information with the full covariance matrix for all charged particles emitted within the angular acceptance of the tracker volume. Only tracks with a minimum of six measurement points are retained. Primary vertices are reconstructed using these tracks as input, based on a simple $\chi^2$ minimization with constraints, yielding three-dimensional vertices with their $\chi^2$ and covariance matrix~\cite{vertexing}. 

\subsection{Reconstruction of electrons and muons}
The identification of electrons and muons in Delphes relies on the true type of the generated particle. True muons and electrons are turned into reconstructed objects if the generated particles have a transverse momentum $p_\text{T}>\SI{0.1}{\giga\electronvolt}$ and are within the acceptance of the tracker volume. The momentum of the reconstructed object is taken from the corresponding track, with a 20\% resolution degradation added for electrons to account for bremsstrahlung in the detector material. 

For this analysis, the reconstructed electrons and muons from Delphes are further required to be isolated from other objects in the event. The isolation is based on the sum of the momenta of all charged and neutral reconstructed particles within a $\Delta R=0.4$ cone around the lepton. Leptons where the ratio of this sum and the momentum of the lepton itself is smaller than 0.12 (electrons) or 0.25 (muons) are retained for analysis.

\subsection{Secondary vertex reconstruction}
The reconstruction of DVs is performed within FCCAnalyses using the Secondary Vertex Finder of the LCFIPlus framework \cite{Suehara:2015ura} and the tracks from Delphes. The default algorithm is optimized for heavy-flavor SM decays and has to be slightly adjusted to efficiently reconstruct the larger displacements and masses of the long-lived scalars. 
The algorithm uses all tracks which are not associated to the primary vertex and starts by building all possible track pairs as seeds to the secondary vertex finder. A series of selections are applied to the two-track vertices to improve the quality. These selections include an upper limit on the invariant mass of the track pair, set to \SI{40}{\giga\electronvolt}, as well as an upper limit on the $\chi^2$ value of the vertex, set to 9. Additional tracks are attached to the seed vertices if the addition brings a $\chi^2$ contribution to the vertex smaller than 5. After every addition of a new track, the vertex is refitted and subject to the same requirements on vertex mass and $\chi^2$ as the two-track pairs.

In order to reduce the rate of background DVs, this analysis employs a custom track selection in addition to the selection of non-primary tracks to seed the vertexing. 
Tracks are required to have a transverse momentum $p_\text{T} > \SI{1}{\giga\electronvolt}$ and transverse impact parameter $|d_0| > \SI{2}{\milli\meter}$ to be considered for vertexing. DVs are only reconstructed up to a radius of \SI{2}{\meter}, corresponding to the radial acceptance of the tracking system of the IDEA detector.

\begin{figure}[ht]\centering
  \subfloat[]{\includegraphics[width=0.45\textwidth]{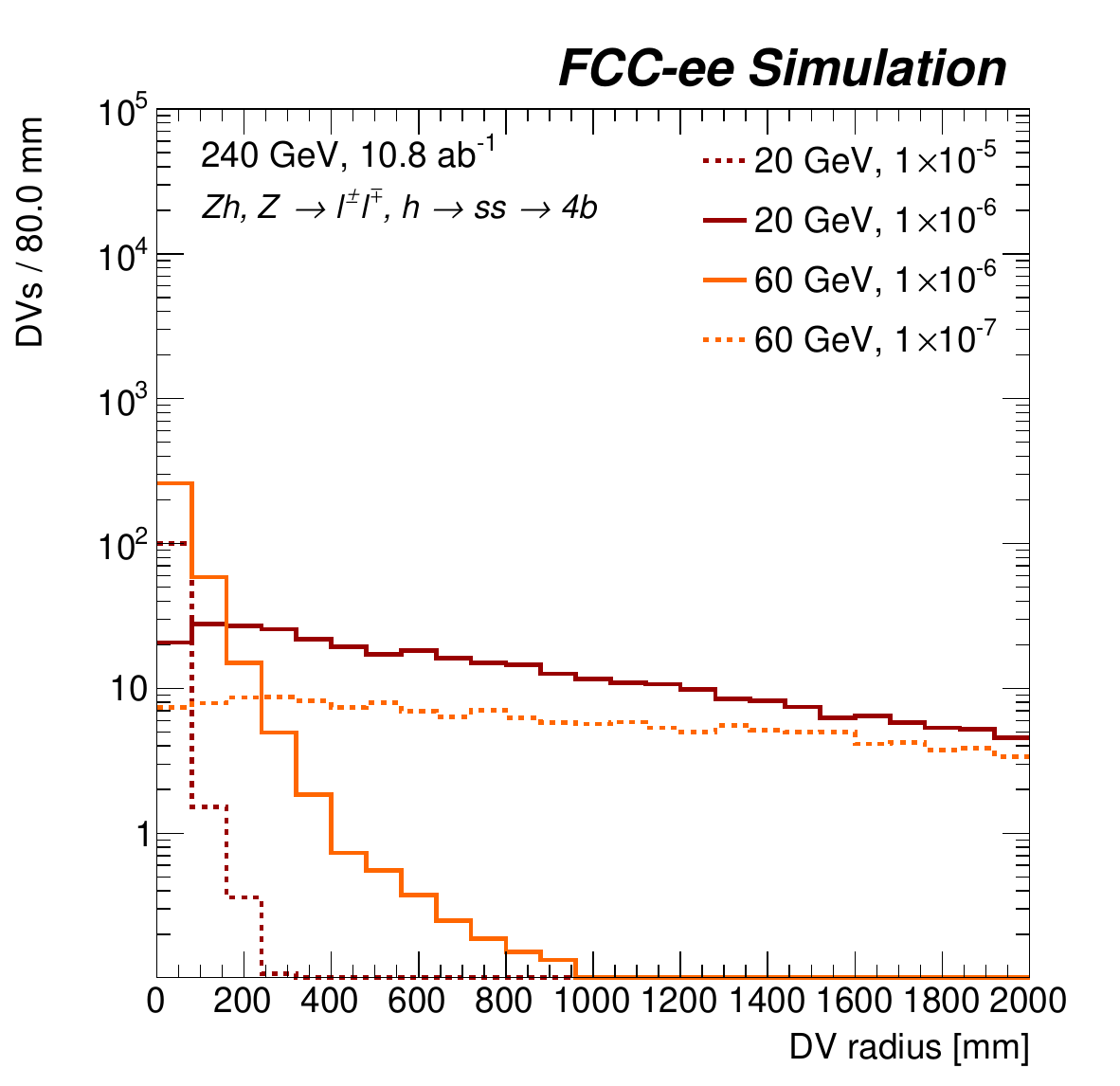} \label{fig:DV_radius_full}}
  \subfloat[]{\includegraphics[width=0.45\textwidth]{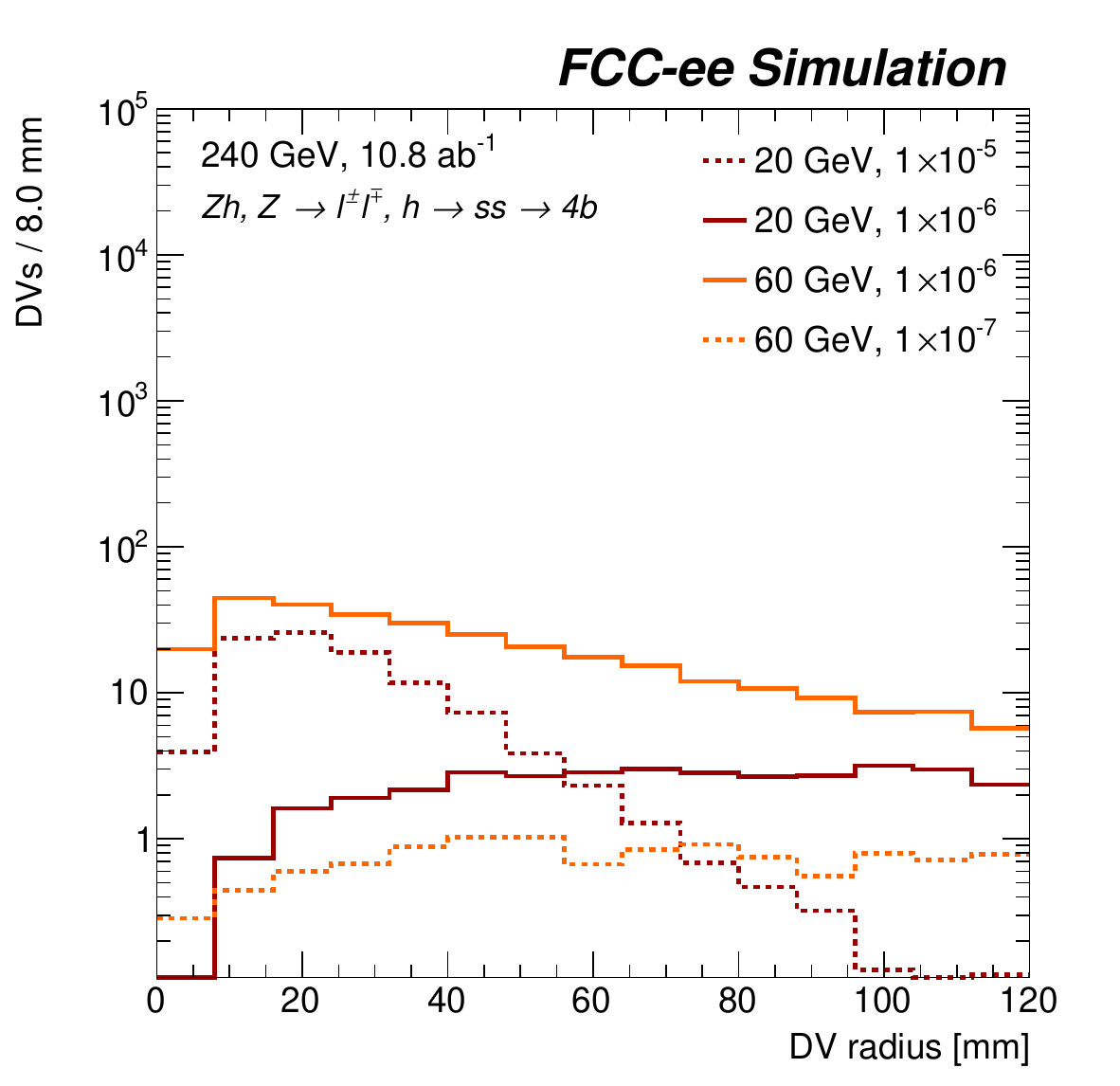} \label{fig:DV_radius_zoom}}
 \caption{Radial distribution of reconstructed DVs on \protect\subref{fig:DV_radius_full} the full scale of the tracking system of the IDEA detector and \protect\subref{fig:DV_radius_zoom} zoomed into the smallest radial region.}
\label{fig:DV_radius}
\end{figure}

Figure~\ref{fig:DV_radius} shows the radial distribution of reconstructed DVs on the scale of the IDEA tracking volume and zoomed in on the smallest radial region, for the same signal points as shown in figure~\ref{fig:genlevel}. Since the analysis targets dark scalar decays into $b\bar{b}$, a single dark scalar can give rise to multiple DVs, due to the $B$-hadron lifetime and subsequent decay chain. Although the algorithm was not optimized to separately reconstruct such nearby vertices, in some cases this will occur, resulting in events with more than two DVs. The radial distribution of the DVs is however quantitatively in agreement with the distribution of the decay length of the dark scalar shown in figure~\ref{fig:decaylength}. In the smallest radial region, around $r<\SI{10}{\milli\meter}$, the distribution of reconstructed DVs is shaped by the track requirements which reduce the efficiency. Below a radius $r=\SI{2}{\milli\meter}$, the reconstruction efficiency is zero due to the $|d_0|$ requirement on the seed tracks. While this affects the signal samples with the shortest lifetimes, it is essential for the reduction of background DVs from heavy-flavor decays.

\section{Event selection}~\label{section:eventselection}
\begin{figure}[ht]\centering
  \subfloat[]{\includegraphics[width=0.45\textwidth]{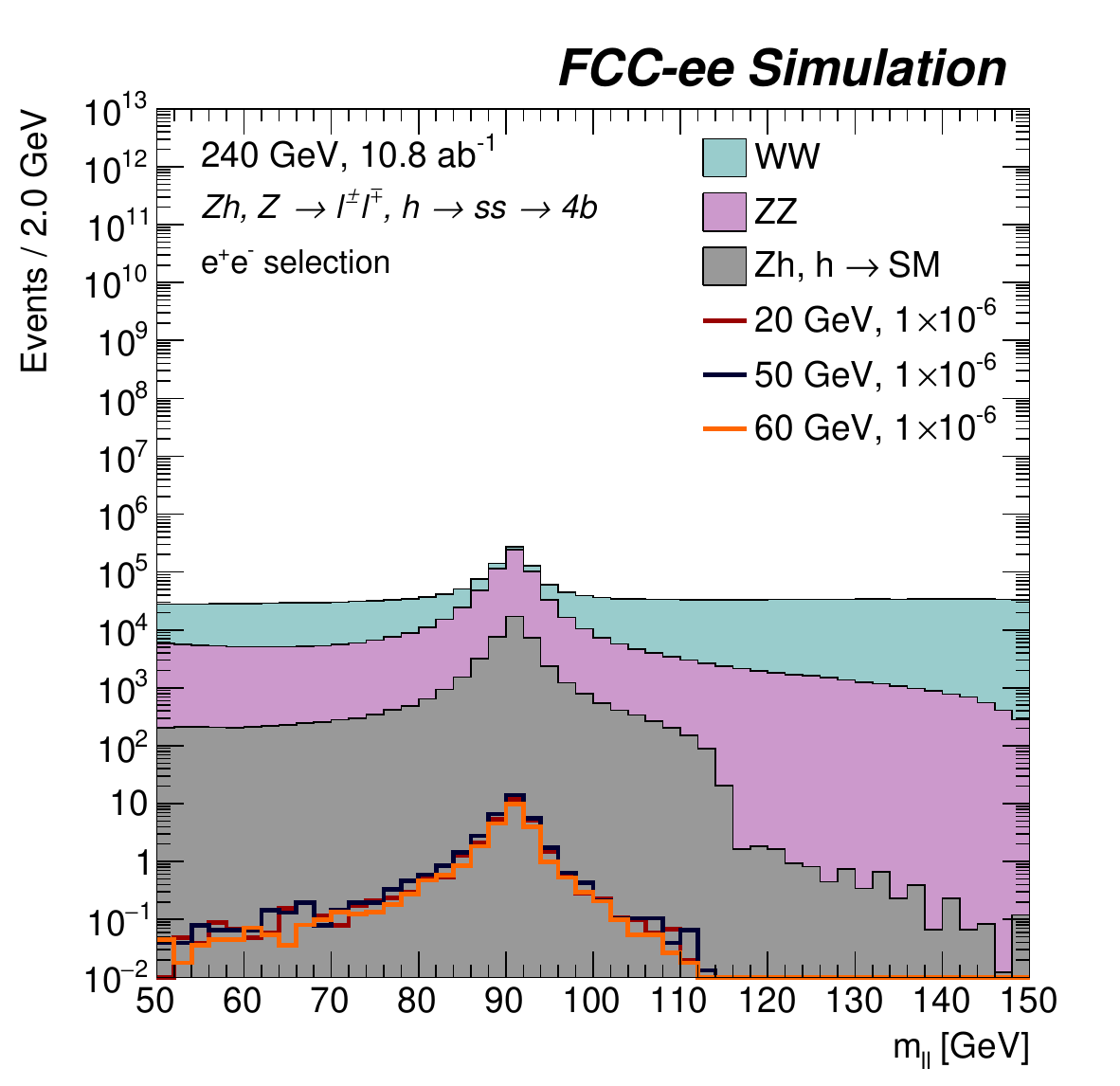} \label{fig:mll_ee}}
  \subfloat[]{\includegraphics[width=0.45\textwidth]{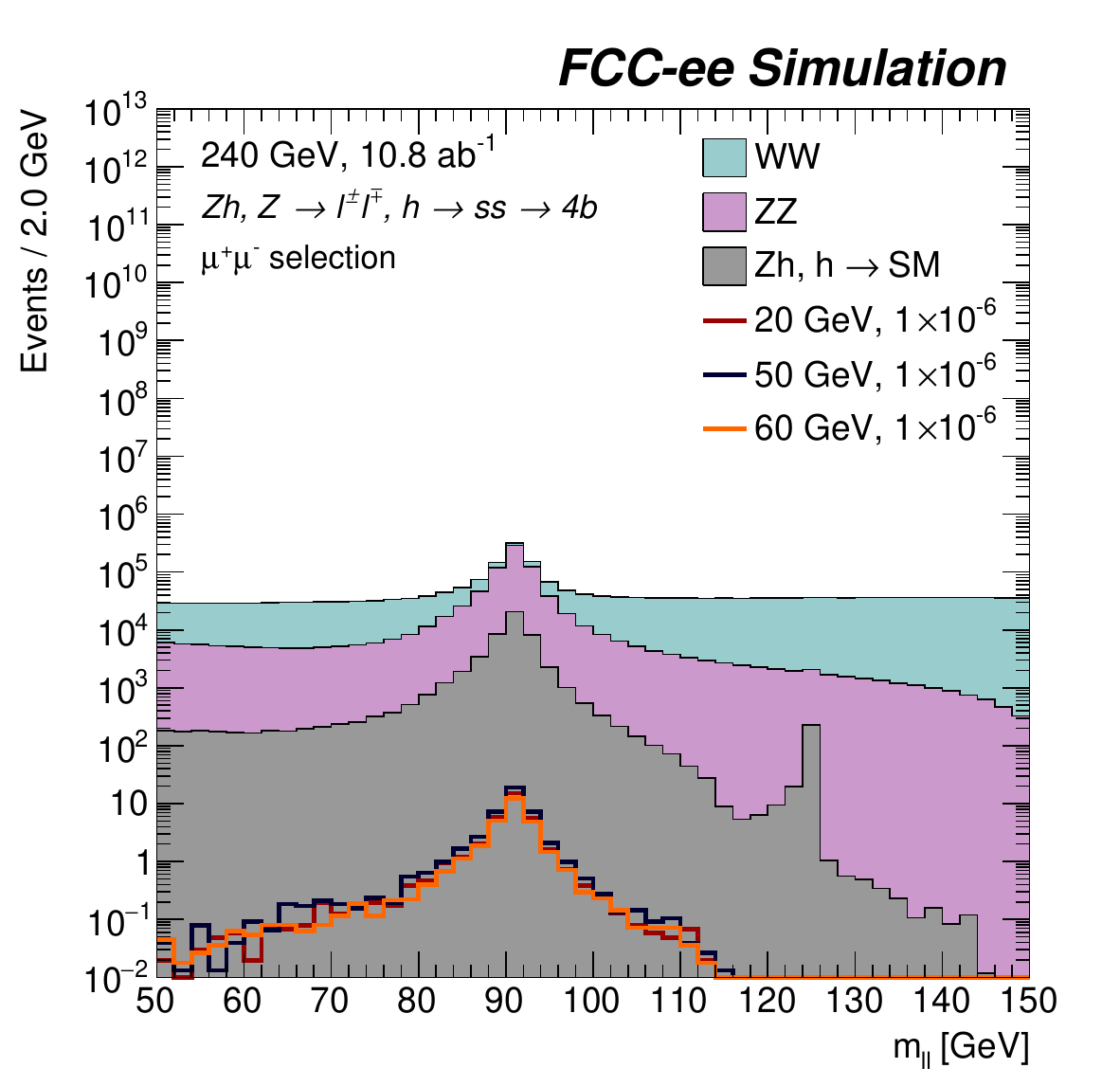}\label{fig:mll_mm}}
 \caption{Reconstructed dilepton invariant mass in events with exactly two opposite-charge \protect\subref{fig:mll_ee} electrons or \protect\subref{fig:mll_mm} muons.}
\label{fig:mll}
\end{figure}

The sensitivity to the long-lived dark scalar decays at the FCC-ee is evaluated using an event selection designed to achieve a background-free search. Events are considered for analysis if they contain two opposite-charge electrons or muons and no additional leptons. Figure~\ref{fig:mll} shows the invariant mass of the lepton pair for $e^+e^-$ and $\mu^+\mu^-$ events in background and a few example signal samples. Although not visible on the scale of the figures, the electron-pair spectrum is slightly wider than the muon-pair spectrum, reflecting on the lower electron resolution simulated in Delphes. Only events where the dilepton invariant mass is in the range $\SI{70}{\giga\electronvolt} < m_{ll} < \SI{110}{\giga\electronvolt}$ are retained for analysis, in order to suppress background from $WW$ production. Events passing these lepton requirements are referred to as pre-selected events.

\begin{figure}[ht]\centering
  \subfloat[]{\includegraphics[width=0.45\textwidth]{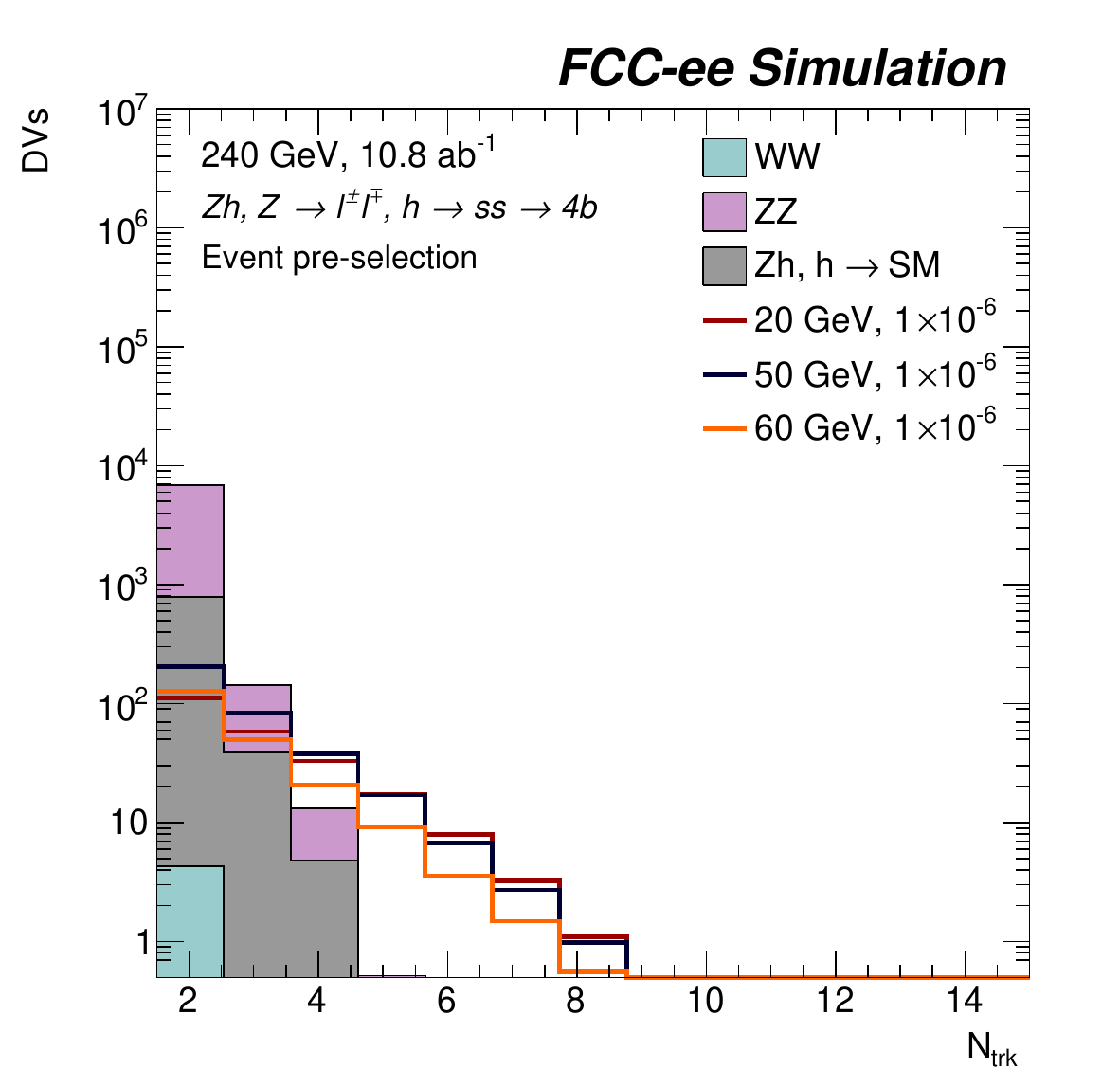} \label{fig:DV_n_trk}}
  \subfloat[]{\includegraphics[width=0.45\textwidth]{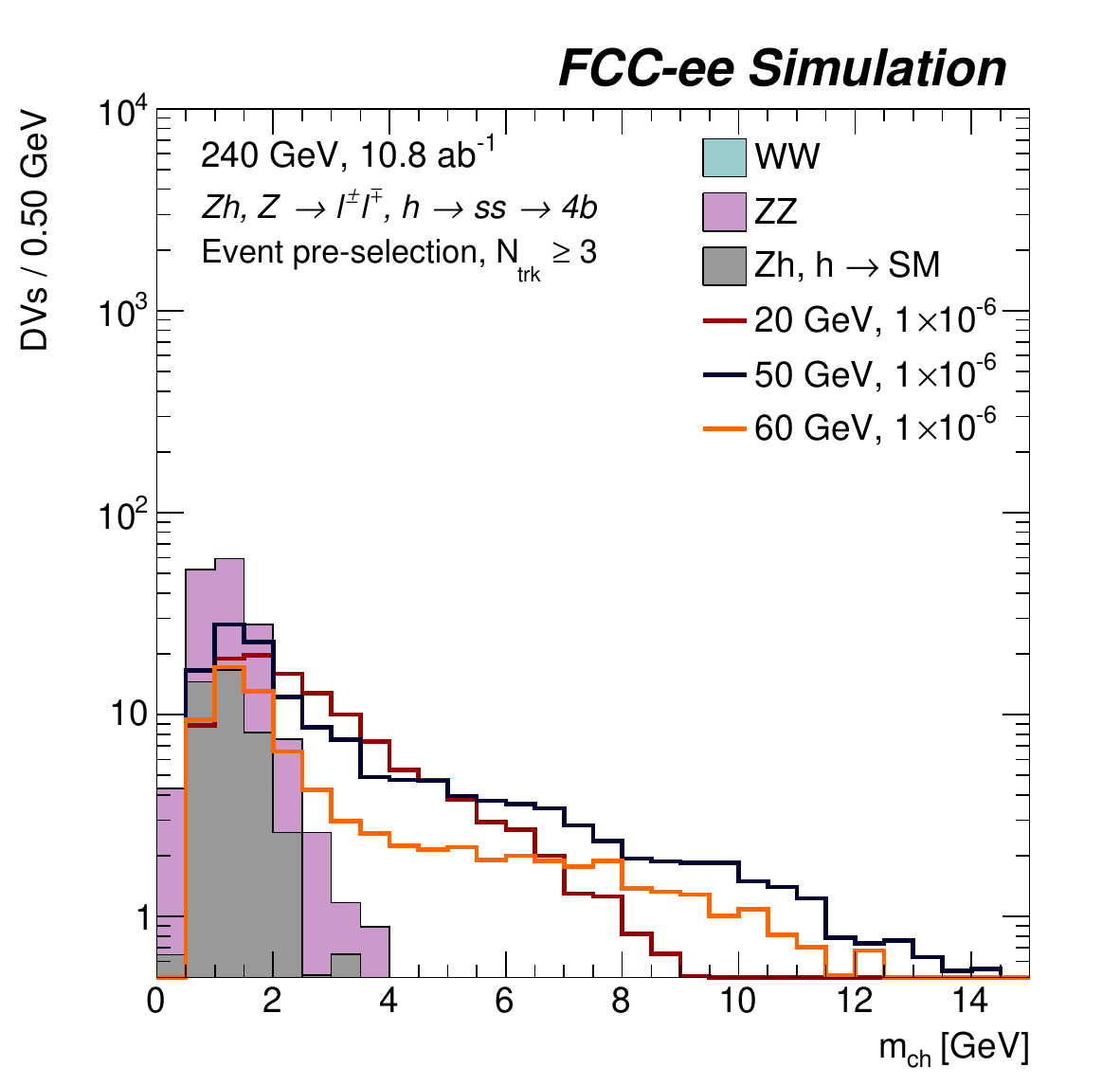}\label{fig:DV_mass}}
 \caption{\protect\subref{fig:DV_n_trk} DV track multiplicity and \protect\subref{fig:DV_mass} DV charged invariant mass for DVs with $N_\text{trk}\geq3$ in pre-selected events.}
\label{fig:DV_properties}
\end{figure}

After pre-selection, the DVs in the events are subject to a series of requirements designed to reduce the number of DVs in background. Figure~\ref{fig:DV_n_trk} shows the number of tracks, $N_\text{trk}$, in each DV for background and a few example signal samples. DVs are required to have at least three tracks to pass the DV selection. Figure~\ref{fig:DV_mass} shows the charged invariant mass, $m_\text{ch}$, computed from all tracks in DVs with three or more tracks, assigning the pion mass to each track. The DV selection requires the charged invariant mass to be $m_\text{ch}>\SI{2}{\giga\electronvolt}$.

\begin{figure}[ht]\centering
\includegraphics[width=0.7\textwidth]{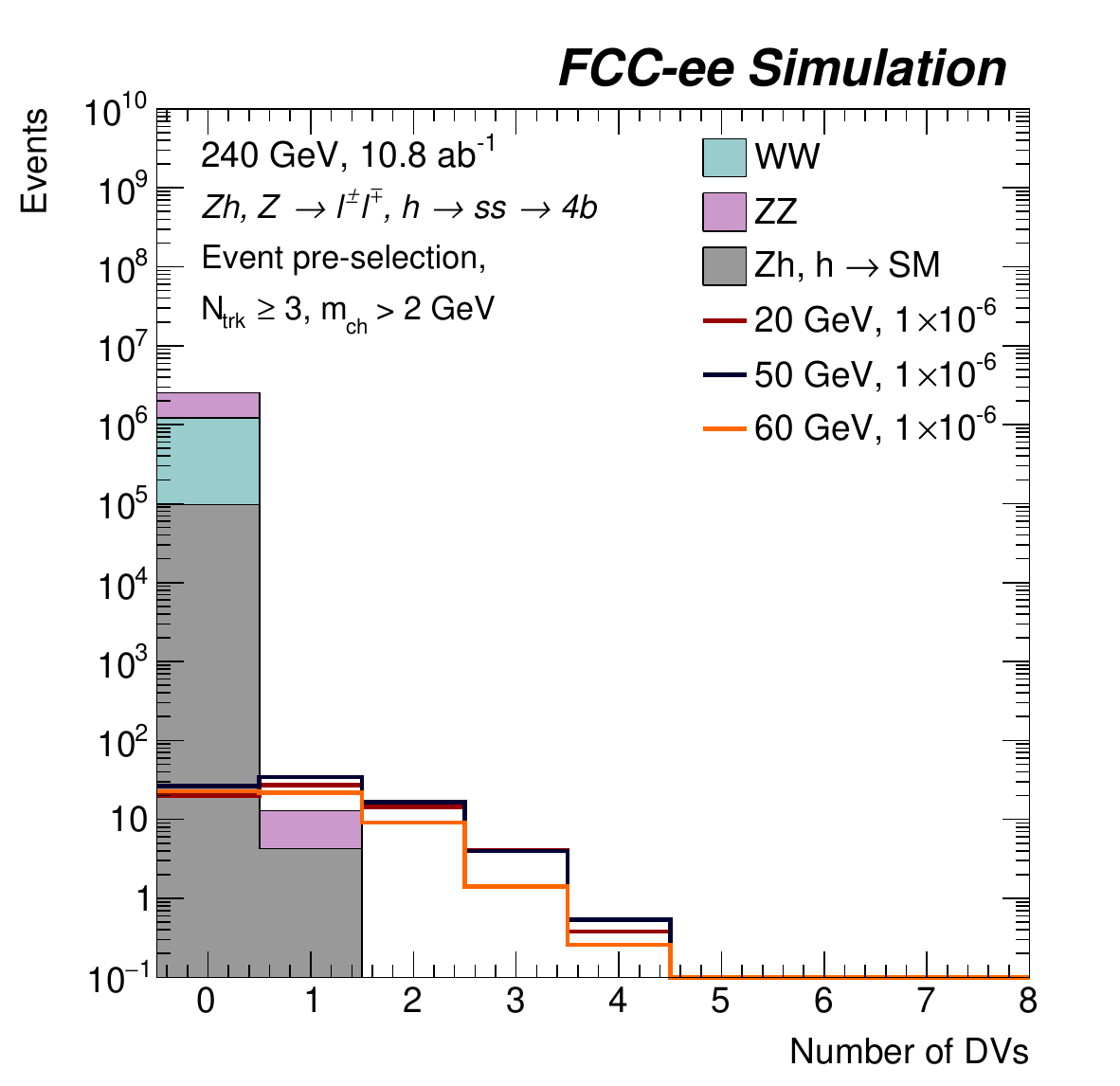}
    \caption{Number of DVs passing the full DV selection in pre-selected events.}
    \label{fig:nDVs}    
\end{figure}

Pre-selected events are eventually selected for analysis based on the number of DVs passing the above requirements. This quantity is shown in figure~\ref{fig:nDVs}. The final event selection requires each event to contain at least two DVs, which efficiently removes all SM background. All event and DV selections are summarized in table~\ref{tab:evtsel}.
The number of events before any selection, after pre-selection and after final selection, are shown in table~\ref{tab:cutflow} for each background process and three example signal samples. While the number of background events is zero after final selection, this number has to be considered with the statistical uncertainty on the number of pre-selected events. The study would therefore benefit from larger MC statistics for a final confirmation of the background-free nature of the analysis.

The acceptance times efficiency for the selection, defined as the number of selected events divided by the number of generated events, ranges between zero and \SI{32}{\percent} for the signal. The highest number is reached for the (\SI{50}{\giga\electronvolt}, \num{3e-7}) sample for which \SI{99}{\percent} of the dark scalars decay within the radial region $\SI{2}{\milli\meter}<r<\SI{2}{\meter}$ where DVs can be reconstructed. 
For the (\SI{60}{\giga\electronvolt}, \num{1e-5}) sample, which has the shortest mean lifetime, all dark scalar decays fall within the $r<\SI{2}{\milli\meter}$ region yielding zero acceptance. Similarly, for the (\SI{20}{\giga\electronvolt}, \num{1e-7}) signal point, which has the longest mean lifetime, \SI{94}{\percent} of the dark scalars decay outside the tracking system, and the acceptance times efficiency reaches \SI{0.4}{\percent}.

\begin{table}[ht]
\centering
\resizebox{0.9\textwidth}{!}{%
\begin{tabular}{@{}lll@{}}
\toprule
\multicolumn{1}{l|}{\multirow{3}{*}{Event pre-selection}} & Lepton content & One $e^+e^-$ or $\mu^+\mu^-$ pair \\
\multicolumn{1}{l|}{} & & and no additional leptons                                \\
\multicolumn{1}{l|}{} & Di-lepton invariant mass & $\SI{70}{\giga\electronvolt} < m_{ll} < \SI{110}{\giga\electronvolt}$ \\ \midrule
\multicolumn{1}{l|}{\multirow{2}{*}{DV selection}} & Track multiplicity  & $N_\text{trk}\geq3$  \\
\multicolumn{1}{l|}{} & Charged invariant mass & $m_\text{ch}>\SI{2}{\giga\electronvolt}$ \\ \midrule
\multicolumn{1}{l|}{Final event selection} & DV multiplicity & $N_\text{DV}\geq2$ \\ \bottomrule
\end{tabular}
}
\caption{Summary of the requirements defining the event pre-selection, DV selection, and final event selection.}
\label{tab:evtsel}
\end{table}

\begin{table}[ht]
\setlength\tabcolsep{12pt} 
\centering
\resizebox{0.9\textwidth}{!}{%
\begin{tabular}{@{}lr@{\hspace*{2pt}}rr@{\hspace*{2pt}}rr@{}}
\toprule
Process & \multicolumn{2}{r}{No selection} & \multicolumn{2}{r}{Pre-selection} & Final selection \\ \midrule \midrule
$Zh$ & \num{1398880} $\pm$ & \hspace{1pt}\num{970}  & \num{97291} $\pm$ &\num{32} & \multicolumn{1}{c}{\hspace{4pt}0}  \\ 
$ZZ$ & \num{14677100} $\pm$ & \hspace{1pt}\num{1000}  & \num{1323650} $\pm$ & \num{300} & \multicolumn{1}{c}{\hspace{4pt}0} \\ 
$WW$ & \num{177535800} $\pm$ & \hspace{1pt}\num{6300}  & \num{1119160} $\pm$ & \num{500} & \multicolumn{1}{c}{\hspace{4pt}0}   \\ \hline 
Total Background & \num{193611800} $\pm$ & \hspace{1pt}\num{6500} & \num{2540100} $\pm$ & \num{590} & \multicolumn{1}{c}{\hspace{4pt}0} \\ \midrule
$\SI{20}{\giga\electronvolt}, \num{1e-6}$ & \num{97.46} $\pm$ & \num{0.10}  & \num{66.30} $\pm$ & \num{0.08} & \num{18.92} $\pm$ \num{0.04} \\ 

$\SI{50}{\giga\electronvolt}, \num{1e-6}$ & \num{130.92} $\pm$ & \num{0.15}  & \num{81.83} $\pm$ & \num{0.12} & \num{21.02} $\pm$ \num{0.06} \\ 

$\SI{60}{\giga\electronvolt}, \num{1e-6}$ & \num{89.17} $\pm$ & \num{0.08}  & \num{55.27} $\pm$  & \num{0.07}   & \num{10.77} $\pm$  \num{0.03} \\ \bottomrule 

\end{tabular}
}
\caption{Number of events with no selection, after pre-selection, and final selection for the background processes and three example signal samples.}
\label{tab:cutflow}
\end{table}

\section{Results}~\label{section:results}
\begin{figure}[ht]\centering
  \subfloat[]{\includegraphics[width=0.5\textwidth]{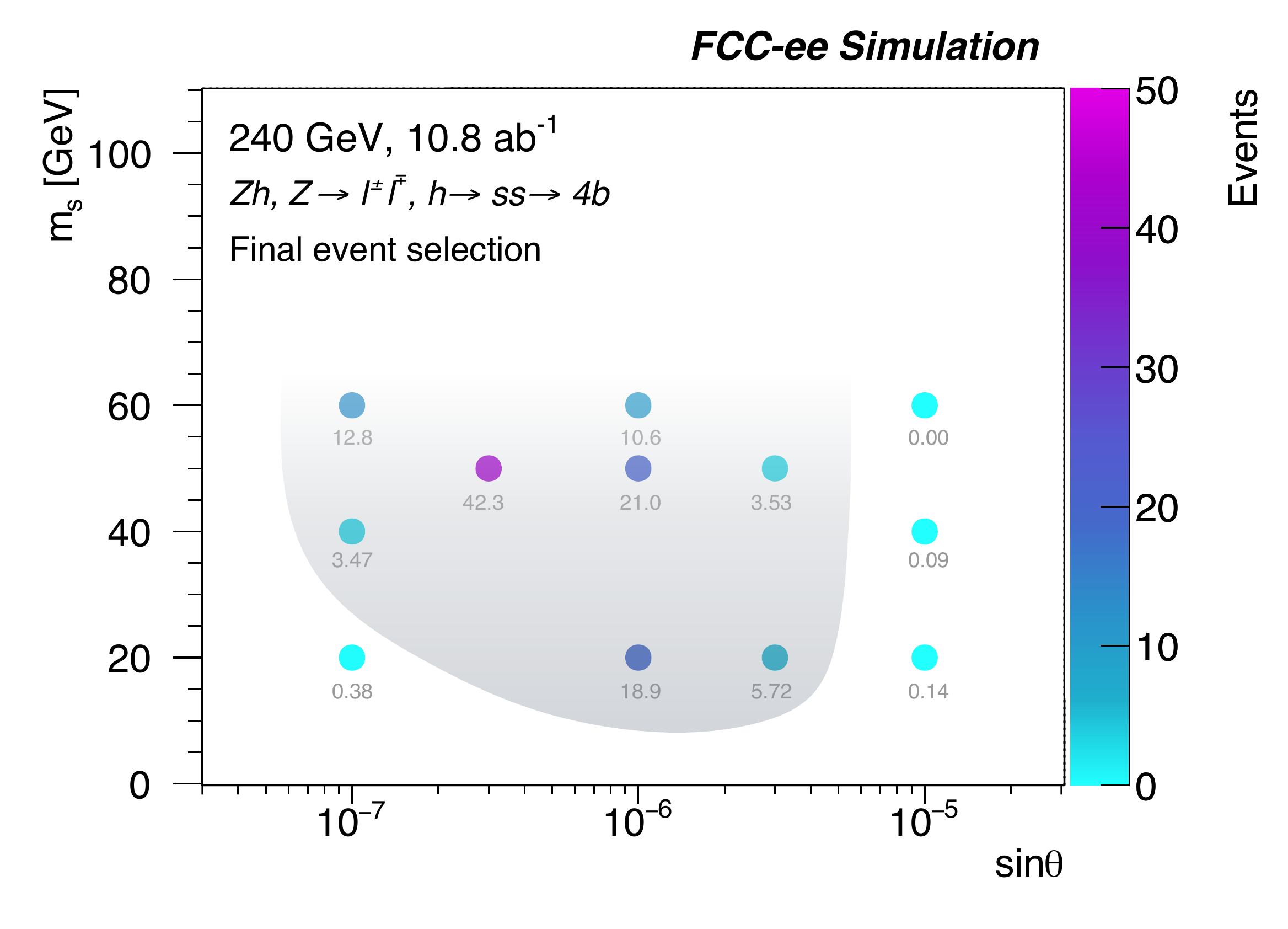} \label{fig:sensitivity_msintheta}}
  \subfloat[]{\includegraphics[width=0.5\textwidth]{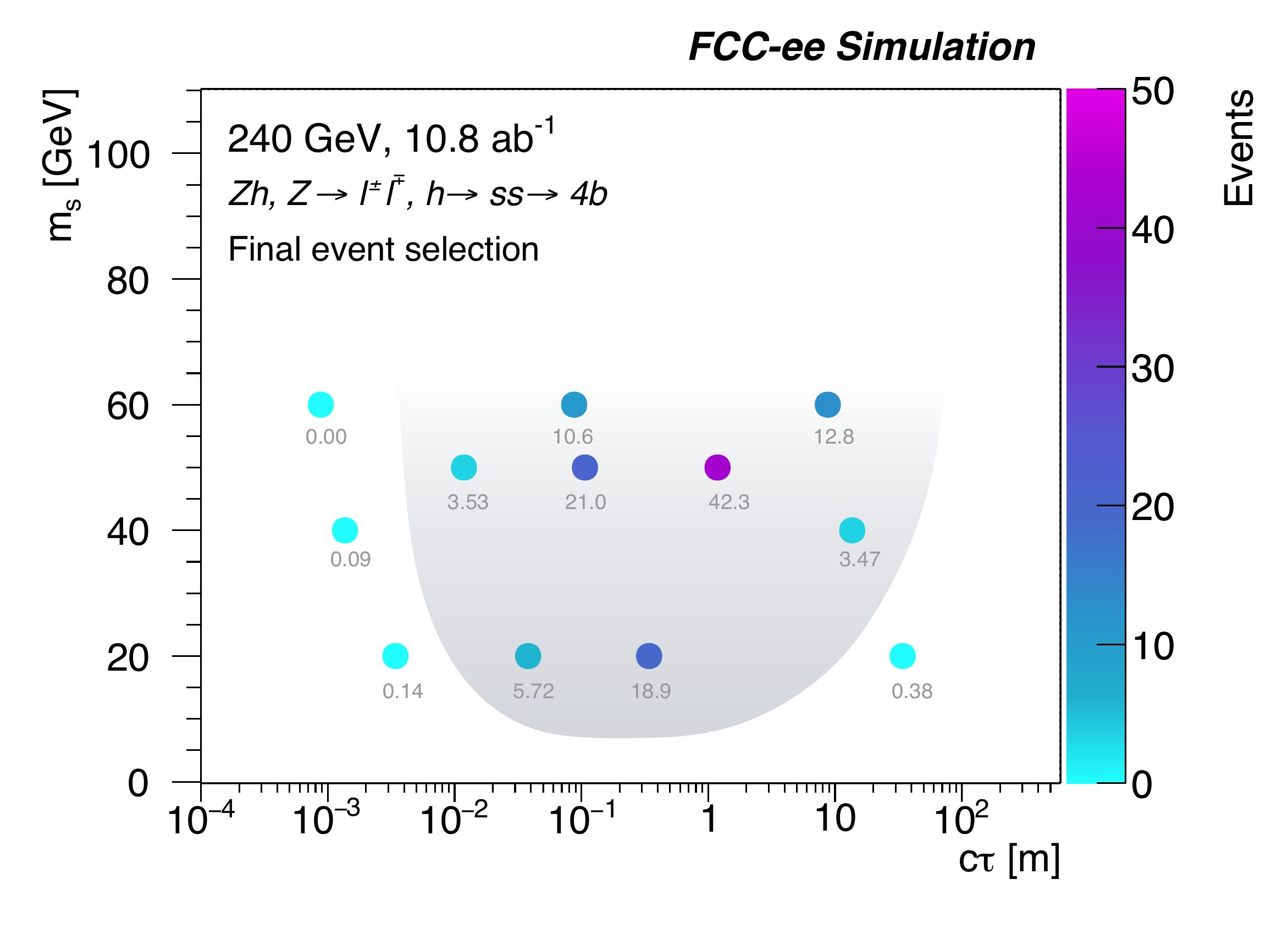} \label{fig:sensitivity_ctaubr}}
 \caption{Number of selected signal events as a function of  \protect\subref{fig:sensitivity_msintheta} dark scalar mass $m_s$ and mixing parameter $\sin\theta$ and \protect\subref{fig:sensitivity_ctaubr} dark scalar mass $m_s$ and lifetime $c\tau$. The shaded region is drawn around signal points with at least three events.}
\label{fig:sensitivity}
\end{figure}

The event selection reduces the background to zero and discovery is therefore possible for any signal point with at least three selected events. The number of events passing the final event selection is shown for all signal samples in figure~\ref{fig:sensitivity_msintheta} for the parameter space spanned by the dark scalar mass and mixing angle, and in figure~\ref{fig:sensitivity_ctaubr} for the parameter space spanned by the dark scalar mass and lifetime $c\tau$. 
A shaded region is drawn around signal points with at least three events to indicate an approximate region of sensitivity to the signal. The results show that the search provides sensitivity to dark scalar masses between $m_s=\SI{20}{\giga\electronvolt}$ and $m_s=\SI{60}{\giga\electronvolt}$ and mean proper lifetimes $c\tau$ between approximately \SI{10}{\milli\meter} and \SI{10}{\meter} for the simulated Higgs to dark scalar branching ratios around \SI{0.1}{\percent}.

While the simulation is performed for a coupling parameter $\kappa=0.0007$, the results can be rescaled to lower values, corresponding to lower Higgs to dark scalar branching ratios. The highest number of selected events is achieved for the (\SI{50}{\giga\electronvolt}, \num{3e-7}) signal point which rescaled to $\kappa=0.0002$ has a branching ratio at \num{9e-5} and 3.3 selected events. This exercise suggests that the search for long-lived dark scalars at the FCC-ee with the IDEA detector concept has potential to probe $h\to ss$ branching ratios below \num{e-4} in a lifetime regime around $c\tau=\SI{1}{\meter}$, providing complementary sensitivity to the HL-LHC.

\section{Concluding remarks}~\label{section:conclusions}
This paper presents the first sensitivity analysis for exotic Higgs boson decays at the electron-positron stage of the FCC-ee within the FCCAnalyses framework. The work considers the production of $Zh$ events at $\sqrt{s}=\SI{240}{\giga\electronvolt}$, with the $Z$ boson decaying leptonically and the Higgs boson decaying into two long-lived dark scalars which further decay into $b\bar{b}$. The event selection requires the presence of an opposite-charge electron or muon pair with invariant mass consistent with the $Z$ boson, and at least two DVs in the final state. This is seen to remove the SM background, while retaining sensitivity to dark scalar masses between $m_s=\SI{20}{\giga\electronvolt}$ and $m_s=\SI{60}{\giga\electronvolt}$ and mean proper lifetimes $c\tau$ between \SI{10}{\milli\meter} and \SI{10}{\meter}. The results suggest that the search strategy has potential to probe Higgs to dark scalar branching ratios as low as \num{e-4} for dark scalar lifetimes around \SI{1}{\meter}. 


\acknowledgments
We would like to thank Juliette Alimena for useful discussions on long-lived particle searches at FCC-ee, Patrizia Azzi, Emmanuel Perez and Michele Selvaggi for guidance on the FCCAnalyses software, for carefully reading the manuscript and for useful comments,  and finally Kunal Gautam for providing support on the secondary vertexing algorithm in FCCAnalyses.
G.~Ripellino is supported by the Carl Trygger foundation (CTS 20:1169). The Swedish Research Council supports M.~Vande Voorde (VR 2018-00482), and A.~Gall\'en and R.~Gonzalez Suarez (VR 2023-03403).

\bibliographystyle{JHEP}
\bibliography{main.bib}

\end{document}